\documentclass{bioinfo}
\copyrightyear{2015}
\pubyear{2015}

\begin{document}
\firstpage{1}

\title[TAG for metatranscriptome analysis]{Utilizing de Bruijn graph of metagenome assembly for metatranscriptome analysis}
\author[Yuzhen Ye and Haixu Tang]{Yuzhen Ye\,$^1$\footnote{To whom correspondence should be addressed.}\,  and Haixu Tang$^1$}
\address{$^{1}$ School of Informatics and Computing, Indiana University\\
 150 S Woodlawn Ave, Bloomington, IN 47405}

\history{Received on XXXXX; revised on XXXXX; accepted on XXXXX}

\editor{Associate Editor: XXXXXXX}

\maketitle

\begin{abstract}

\section{Motivation:}
Metagenomics research has accelerated the studies of microbial organisms, providing insights into the composition and potential functionality of various microbial communities. Metatranscriptomics (studies of the transcripts from a mixture of microbial species) and other meta-omics approaches hold even greater promise for providing additional insights into functional and regulatory characteristics of the microbial communities. Current metatranscriptomics projects are often carried out without matched metagenomic datasets (of the same microbial communities). For the projects that produce both metatranscriptomic and metagenomic datasets, their analyses are often not integrated. Metagenome assemblies are far from perfect, partially explaining why metagenome assemblies are not used for the analysis of metatranscriptomic datasets.  

\section{Results:}
Here we report a reads mapping algorithm for mapping of short reads onto a de Bruijn graph of assemblies. A hash table of junction k-mers (k-mers spanning branching structures in the de Bruijn graph) is used to facilitate fast mapping of reads to the graph.  We developed an application of this mapping algorithm: a reference based approach to metatranscriptome assembly using graphs of metagenome assembly as the reference. Our results show that this new approach (called TAG) helps to assemble substantially more transcripts that otherwise would have been missed or truncated because of the fragmented nature of the reference metagenome.  

\section{Availability:}
TAG was implemented in C++ and has been tested extensively on the linux platform. It is available for download as open source at http://omics.informatics.indiana.edu/TAG.

\section{Contact:} \href{yye@indiana.edu}{yye@indiana.edu.}
\end{abstract}

\section{Introduction}

Metagenomes are being generated at an accelerating pace, revealing important properties of microbiomes. Other meta-omic (e.g., metatranscriptomic and metaproteomic) techniques can provide additional insights, in particular into functional characteristics of microbial communities, such as gene activities and their regulatory mechanisms. Bacteria have low inventories of short-lived mRNAs; as such, fluctuations in their mRNA pools provide a highly sensitive bioassay for environmental signals (e.g., the concentrations of dissolved organic carbon (\citealp{Shi12}) and pollutant concentrations (\citealp{soil12}) relevant to microbes (\citealp{Moran12})). The acquisition of meta-omics data on human microbiomes will enable us to refine the annotations of the metagenomes (the ENCODE (\citealp{encode12}) and modENCODE (\citealp{modencode10}) projects are great exemplars), and more importantly to study gene activity and its regulation (\citealp{xenobiotics13}) in complex microbial communities in order to understand how microbial organisms work as a community in response to changes in their environment, e.g., health conditions of their human hosts (\citealp{oral14}). A recent metatranscriptomic study of the human oral microbiome using patient-matched healthy and diseased (periodontal) samples revealed that health- and disease-associated communities exhibit defined differences in metabolism that are conserved between patients while the metabolic gene expression of individual species was highly variable between patients (\citealp{oral14}). 

In a metatranscriptomic RNA-seq study, total RNA is first isolated from the sample (with rRNAs removed to enrich for mRNA), which is then reverse transcribed into cDNA, and subjected to sequencing using next-generation sequencing platforms (\citealp{Gosalbes11}). Unlike metagenomics, which reveals potential activity (as reflected in genes or pathways that can be coded for by metagenomic sequences), metatranscriptomic data indicates which of the genes/metabolic pathways are actually active (and the level of their activities) on the basis of their transcription within the community. Giannoukos et al.  presented a protocol for metatranscriptomic analysis of bacterial communities that accommodates both intact and fragmented RNA and combines efficient rRNA removal with strand-specific RNA-seq (\citealp{Giannoukos12}). Currently, only a handful of metatranscriptomic datasets are available (and metaproteomic datasets are even scarcer), but we envision a flood of metatranscriptomic data in the near future, as experimental techniques mature (\citealp{Giannoukos12,curtis14}).  

Metatranscriptome analyses typically include the assignment of the predicted function and taxonomic origin of RNA-seq reads, by directly searching metatranscriptomic sequences (bags of reads) against prokaryotic genomes (the reference genomes) (\citealp{pipeline13}) or known protein sequences (\citealp{curtis14}). This way, tools and pipelines---including MG-RAST (\citealp{mgrast}), MEGAN (\citealp{megan4}) and HUMAnN (\citealp{humann})---that have been developed for metagenome data analysis can be utilized for analyzing metatranscriptomic datasets. For example, Franzosa et al analyzed metagenomic and metatranscriptomic datasets of human gut microbiomes using the HUMAnN pipeline, revealing that metatranscriptional profiles were significantly more individualized than DNA-level functional profiles (\citealp{curtis14}). One potential pitfall of such approaches is that they cannot identify transcripts of new genes, which however may be better annotated using assembly approaches (de novo or reference based). A recent study (\citealp{Microbiome2014}) compared the performances of currently employed transcriptome assemblers---including Trinity (\citealp{trinity}), Oases (\citealp{oases}), Metavelvet (\citealp{metavelvet}) and IDBA-MT (\citealp{idbamt})---and showed that assembly helps to improve the rate of functional annotation for metatranscriptomic datasets.

A matched metagenome can be very helpful for the analysis of metatranscriptomic dataset. Metagenomes are often represented as contigs and scaffolds (although de Bruijn graphs are often the underlying data structure of the assemblers that were used), and are very fragmented, limiting the utilization of metagenome for metatranscriptome analysis. There are pros and cons with the contig (and scaffold) representations of metagenomes. Most existing computational tools for sequence analysis work with linear representations of assemblies, so these tools (or modified versions) can be employed to analyze these representations of metagenomes. However, metagenomes are often very fragmented, and the connections between contigs or scaffolds are not captured in linear representations, which otherwise could be utilized later. For example, after we assembled two metagenomic datasets of stool samples from the Human Microbiome Project (\citealp{hmp12}), the total lengths of scaffolds and contigs ($\geq$ 300 bps) reported by SOAPdenovo2 (\citealp{soap2}) were about 85 and 90 Mbps, respectively, whereas the total length of the edge sequences in the de Bruijn graph from the same assembly was ~150 Mbps for each. This comparison indicates that the de Bruijn graph representation of the assembly contains ~50\% more sequences than scaffolds reported from the assembler: most of these extra sequences are relatively fragmented sequences connecting long contigs. Furthermore, many short contigs contain only gene fragments; even long contigs contain broken genes at their ends due to the complexity of metagenome assembly (\citealp{genestitch}).

Here we propose a novel application of de Bruijn graphs for metatranscriptomic data analyses, taking advantage of the fact that de Bruijn graph representations of metagenome assemblies contain more information than the contigs and scaffolds reported by assemblers. The de Bruijn graph was first proposed for de novo genome assembly in EULER, replacing the traversal of Hamiltonian paths in the overlap graph by the traversal of Eulerian paths (\citealp{euler01}), and is now employed as an efficient data structure in most short-read assemblers (e.g., Velvet (\citealp{velvet}), ALLPATHS-LG (\citealp{allpath}), SOAPdenovo (\citealp{soap}), and IDBA-UD (\citealp{idbaud})) for single genomes and metagenomes. Our approaches based on de Bruijn graph representation of metagenomes provide a natural way of compressing the data, and, more importantly, allow direct utilization of the graphs. We note that we have developed several applications previously, based on de Bruijn graph representation of genomes and metagenomes, for mining of functional elements (\citealp{integron}) and reads mapping (\citealp{Wang12}), demonstrating the utility of direct computation on de Bruijn graphs. Application of our method to simulated and real metatranscriptomic datasets showed that our approach can significantly improve the assembly of metatranscriptomic datasets, resulting in substantially more transcripts that otherwise would have been missed or truncated because of the fragmented nature of the reference metagenome.   

\section{Methods}
In this paper, we propose a novel algorithm (i.e., {\em read2graph}) for aligning short reads from RNA-seq experiments to de Bruijn graphs of assemblies. We note in this paper we focused on de Bruijn graphs of metagenome assemblies, but the mapping algorithm can be applied to mapping short reads to any de Bruijn graph of assembly.  We also developed an application of the mapping algorithm for metatranscriptome assembly using matched metagenomes as the reference. Based on reads mapping results, we will derive putative transcripts (encoding a single bacterial gene or multiple genes within an operon), using paired-end RNA-seq reads to traverse the de Bruijn graph. We named our transcript assembly approach TAG, in which TA stands for Transcript Assembly, and G is used to emphasize the fact that our approach utilizes the graph of metagenome assembly instead of the linear sequences. We note that our method is different from the de novo approaches to transcriptome assembly, including Trinity (\citealp{trinity}), IDBA-MT (\citealp{idbamt}) (and also a hybrid approach (\citealp{idbamtp}) that utilizes known protein sequences), and that it is different from the traditional reference based assembly approaches. In our method, metatranscriptomic sequences are mapped onto matched metagenomes represented as de Bruijn graphs. So our method represents a new variant of the reference based approaches, which uses the de Bruijn graph of matched metagenome, instead of a genome (or a collection of genomes), as the reference. 

\begin{methods}
\subsection{Fast reads mapping onto de Bruijn graph using a hash table of k-mers spanning branching structure in the graph}
Given a de Bruijn graph, in which each edge represents an assembled unique sequence from metagenomic reads, and a set of short reads from an RNA-seq experiment, the goal of our read2graph algorithm is to find the location of each read on the graph. Because bacterial genes do not have split gene structure, we can assume each read should be contained in the graph as a whole; equivalently, each read, if its location in the graph is known, can be represented as a path (i.e., sequence of edges) in the graph. The reads, therefore, can be classified into two groups depending on the path length: some reads are located within a single edge, whereas many others may cross one or more vertices in the graph. The first class of reads can be mapped to the graph using conventional fast reads mapping algorithms by using all edge sequences longer than the read length as the target sequences. In this paper, we used Bowtie 2 (\citealp{bowtie2}) for this purpose; but other mapping algorithms including BWA (\citealp{bwa}) can be used.  Here we focus on the methods for mapping reads spanning multiple edges (i.e., \emph{multi-edge-spanning reads}; see Figure 1), which cannot be mapped using conventional mapping algorithms. A substantial number of reads may belong to this class, due to the incompleteness of metagenome assembly. 
 
Recall that each vertex in the de Bruijn graph represents a k-mer in metagenomic reads (typically k=23--31 for metagenome assembly (\citealp{gut10,hmp12})). Therefore, as illustrated in Figure 1, each multi-edge-spanning read contains one or more junction k-mers (i.e., corresponding to vertices with either indegree or outdegree $>$ 1): reads A and D span three edges in the graph, and thus each contains two such k-mers, whereas reads B and C span two edges, and thus each contains one such k-mer (Figure 2a). Hence, we can build a hash table for all \emph{junction k-mers} that span branching structures in the de Bruijn graph assembly and then search for their exact occurrences in each putative multi-edge-spanning read (i.e., those that cannot be mapped to the edge sequences) with the assistance of the hash table (Figure 1b). Because each k-mer in the de Bruijn graph is unique (\citealp{euler01}), every k-mer in a read matches at most one k-mer stored in the hash table. Each matched k-mer determines a unique putative location of the multi-edge-spanning read in the graph (i.e., a seed match between the read and the graph), and simultaneously breaks the read into two or more segments (Figure 1c). The seed match will then be used to constrain the alignment between the read and the graph, starting from the seed match, going in opposite directions, using a constrained dynamic programming algorithm allowing only a small number of indels and mismatches. The bandwidth for constrained alignment is set to 7 by default for metatranscriptome assembly using matched metagenome as the reference, and this parameter can be changed by users for other purposes. 

\begin{figure}[htbp]
   \centering
   \includegraphics[trim = 0mm 0mm 0mm 0mm, clip, scale=0.4]{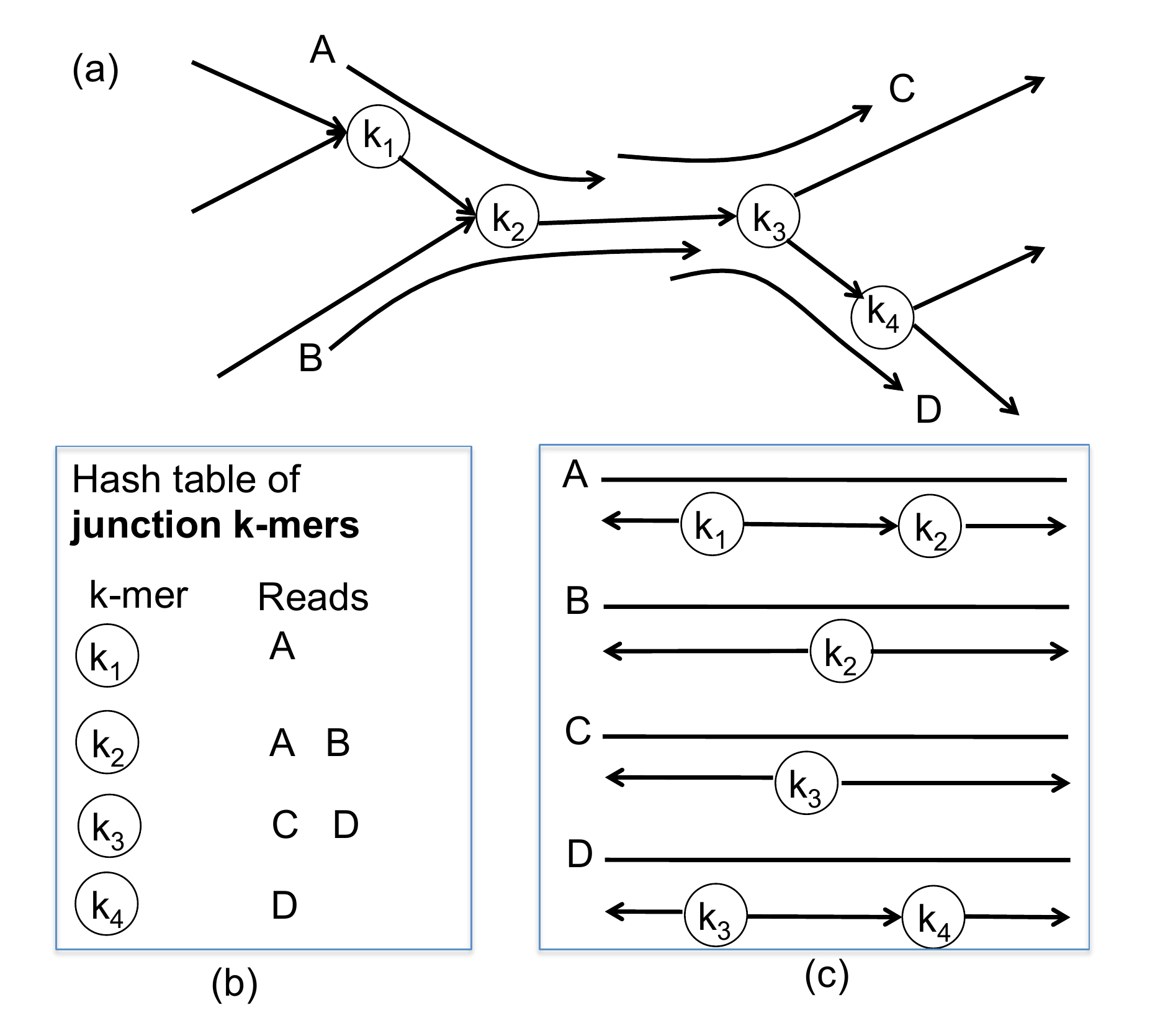}
   \caption{A schematic illustration of the algorithm for mapping reads onto de Bruijn graphs. (a) A toy example showing four reads spanning \emph{junction k-mers} in the graph (shown as the vertices). (b) Using a hash table of junction k-mers, candidates of reads that span multiple edges can be retrieved by looking up in the table. (c) For each candidate, a matched k-mer determines a unique putative location of the read in the graph (i.e., a seed match). The seed match will then be used to constrain the alignment between the read and the graph by a dynamic programming algorithm. }
   \label{fig:hash}
\end{figure}

The mapping of multi-edge-spanning reads should run fast and consume reasonable memory because usually there are only hundreds of thousands of junction k-mers in a typical metagenome assembly in practice. We note that the multi-edge-spanning reads considered here are different from the split reads considered in transcript assembly for eukaryotes (\citealp{trinity}), and in rare cases for archaeal species (due to tRNA splicing and self-splicing introns) (\citealp{splitread13}). Since strand-specific RNA-seq protocols are often used in metatranscriptome analysis (\citealp{Giannoukos12}), our algorithm can consider the strand information and map reads to one appropriate strand in the de Bruijn graph that contains sequences from both DNA strands (and thus is symmetric). 

\begin{figure}[htbp]
   \centering
   \includegraphics[trim = 0mm 0mm 0mm 0mm, clip, scale=0.4]{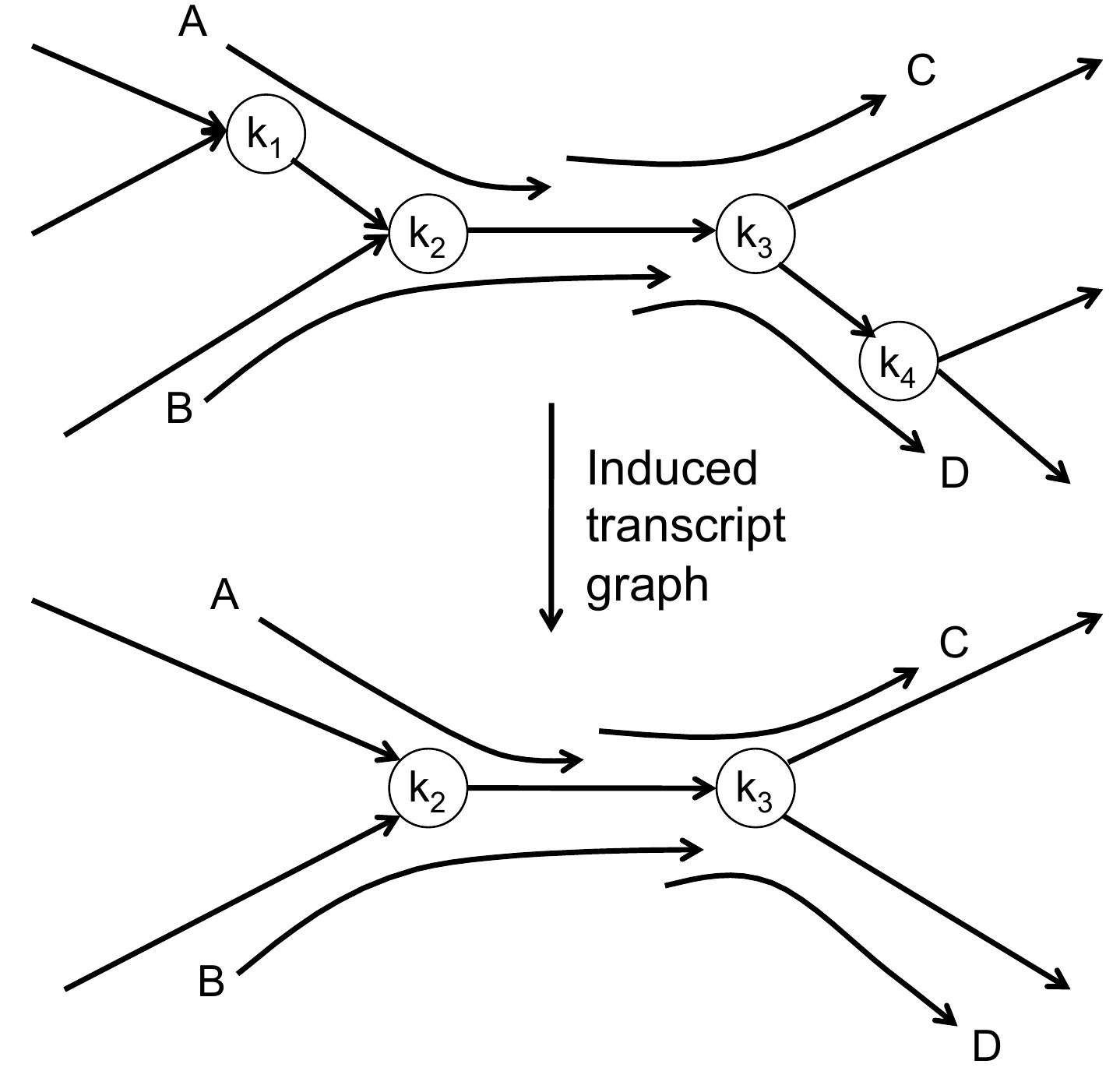}
   \caption{A schematic example illustrating the induced transcript graph derived from four reads (A, B, C and D) mapped to a de Bruijn graph of metagenome assembly. }
   \label{fig:hash}
\end{figure}

\subsection{Construction of transcripts from mapped reads}
Once all RNA-seq reads including multi-edge-spanning reads are mapped to the graph, each read can be represented by a path (referred to as the {\em read path}) traversing the graph $<e_1, e_2, \dots, e_l>$ ($e_1, e_2, \dots, e_l$ are edges; for non-multi-edge-spanning reads, path length $l=1$) as well as two offset values representing the locations of the read in the first and last edges in the graph. Furthermore, in most cases, two paired-end reads can also be represented as a path (i.e., the {\em read-pair path}) if there exists a unique path in the graph whose length is consistent with the expected insert size. As a result, the assembly of RNA-seq reads is equivalent to the superpath problem, which attempts to find a minimal set of superpaths (each corresponding to a transcript) that covers a given set of paths in a de Bruijn graph (\citealp{pop09}). Although this problem is generally hard, we can represent the solutions of the problem in a much simpler subgraph (the transcript graph) that contains only the edges present in at least one of the read paths or read-pair paths. Figure 2 shows such an example: assuming four read paths (A, B, C and D) are derived from multi-edge-spanning reads, we will induce the transcript graph by retaining all edges in these paths, and then contracting all vertices with both indegree and outdegree of 1. We note that many read paths may contract into a single edge in the transcript graph if they are not tangled with reads from another transcript; as a result, the corresponding transcript sequences can be retrieved as a subsequence of an edge sequence in the transcript graph. In other cases, read paths remain spanning multiple edges in the transcript graph. These read paths sometimes can be used to further simplify transcript graph, as illustrated in the heuristic algorithms in genome assembly (\citealp{velvet,euler01}). For instance, in the example shown in Figure 2, if we have two read-pair paths spanning AÐC and BÐD, respectively, we can obtain two resolved transcripts from the graph. Otherwise, we can only obtain partial transcript sequences (see Table 1). We note that, even if the transcripts cannot be fully resolved, the transcript graph is still useful for inferring the abundances of putative transcripts in a metatranscriptome sample based on the counts of reads on the edges in the transcript graph, a problem similar to the inference of splicing variants in eukaryotic RNA-seq experiments (see \citealp{Pachter2011} for a review). 

\subsection{Metatranscriptome assembly using metagenome assembly graph as the reference}
Our approach for metatranscriptome assembly (called TAG) is based on the read2graph mapping algorithm and the transcript construction approach as described above. Given a metatranscriptomic dataset and a matched metagenomic dataset, SOAPdenovo2 (\citealp{soap2}), one commonly used assembler in metagenomic shotgun sequencing, is used to assemble the metagenomic dataset.  Notably, SOAPdenovo2 is a de Bruijn graph-based assembler, and in its final output, both the de Bruijn assembly graph and the contig sequences (representing the {\em edges} in the graph) are produced. The mapping of metatranscriptomic sequences to the de Bruijn graph is conducted in two consecutive steps: 1) all reads are first mapped to the edges (i.e., contigs) in the de Bruijn graph using Bowtie 2 (\citealp{bowtie2}), and then, 2) the un-mapped reads in the previous step are further mapped to the graph based on the matching with junction k-mers. Next, TAG traverses the de Bruijn graph along with the mapped metatranscriptomic reads, and reports the transcripts that may span multiple edges in the assembly graph. We note that other short read assemblers (such as IDBA (\citealp{idbaud})) and mapping tools (such as BWA (\citealp{bwa})) can be utilized for generating the inputs (i.e., the de Bruijn assembly graph and the mapping of metatranscriptomic reads to contigs) for TAG. For the rest of this paper, we will focus on the utility of TAG on improving the assembly of transcripts, which will be demonstrated by using the SOAPdenovo2 and Bowtie2 tools. The construction of an optimal pipeline (in particular the selection of upstream software tools) utilizing TAG is beyond the scope of this paper. 

\end{methods}

\section{Results}

We tested our tool (TAG) on two metatranscriptomic datasets (\citealp{Giannoukos12}): one derived from a mock microbial community consisting of three bacterial species, and the other derived from a real microbiome sample in human stool. Results showed that our graph-based reads mapping algorithm (read2graph) is efficient, and TAG, which is based on the mapping algorithm, significantly improves the assembly of metatranscriptomes by considering reads mapping to branching structures in de Bruijn graphs of matched metagenomes.

\subsection{Evaluation of assembly accuracy on a mock dataset }
We first tested TAG using a metranscriptomic data from the mock bacterial community of three species (\citealp{Giannoukos12}). The `matched' metagenomic dataset used in TAG were simulated from the reference genomes of these bacteria  (\emph{E. coli} [GenBank: NC\_000913.3], \emph{P. marinus} [GenBank: NC\_005072.1] and \emph{R. sphaeroides} [GenBank: NC\_007493.2]) using NeSSM (\citealp{NeSSM}) with the Illumina error model. We used this {\em hybrid} approach here because 1) there is currently no metatranscriptomic dataset from a mock community with a matched metagenomic dataset available, and 2) there is no proper software tool for simulating metatranscriptomic dataset \footnote{Flex Simulator is a tool for simulating RNAseq data for single species, and it has been mainly used for eukaryotic species. Bacteria have complicated transcription regulation mechanisms, which are not completely understood.}. In total, 1M paired-end reads of length 101 bp (i.e., approximated 20X coverage) were simulated from the three species with equal abundances. SOAPdenov2 (k = 31; see below for the choice of k-mer size) was used to assemble the simulated reads, and the assembly results (including the contigs and the de Bruijn assembly graph) were then used as the inputs to TAG. Because this is a simple community with bacterial species that are phylogenetically distant (\citealp{Giannoukos12}),  the assembly graph of the metagenome is not very tangled, and thus we do not anticipate that many transcripts reported by TAG will span multiple edges (referred to as the {\em multi-edge transcripts}) in the assembly graph. In fact, TAG reported a total of 9,428 transcripts (of $>= 100$bp), among which only 138 are multi-edge spanning transcripts. 

\begin{table*}[!t]
\caption{Performance comparison of TAG and other assemblers on the mock dataset.}
\begin{center}
\renewcommand{\arraystretch}{1.4}
\setlength\tabcolsep{3pt}
\begin{tabular}{llll}
\hline
\noalign{\smallskip}
 & Oases & Trinity & TAG\\
 \hline
No. of transcripts$^\#$ & 12,598 & 24,804 $^+$ & 9,428\\
Perfectly aligned transcripts (percentage)$^*$ & 5,483 (43.5\%) & 12,392 (50.0\%) & 9,412 (99.8\%)\\
Transcripts with minor problems (percentage)$^*$ & 2,724 (21.6\%) & 2,725 (11.0\%) & 14 (0.15\%) \\
Problematic transcripts (percentage)$^*$ & 4,391 (34.9\%)$^\&$ & 9,687 (39.1\%)$^\&$ & 2 (0.02\%) \\
Total length of the transcripts & 6,860,841 bps & 7,428,187 bps & 7,020,975 bps\\
Total length of perfectly aligned transcripts & 2,265,224 bps & 3,858,486 bps & 7,002,290 bps\\
Total length of good transcripts & 4,076,481 bps & 5,025,072 bps & 7,020,484 bps\\
 \hline
\end{tabular}
\end{center}
{\footnotesize $^\#$ Only transcripts of at least 100 bps were considered for all programs. $^+$ Trinity has many more transcripts, but their total length is comparable to the other methods. $^*$A transcript that is perfectly aligned to one of the reference genomes (with an alignment covering the entire transcript at 100\% sequence identity) is considered to be correctly assembled. We consider the problem of a transcript is ``minor" if its longest alignment with the reference genomes is not 10 nt shorter than the transcript and the alignment has 95\% sequence identity or better. Other transcripts that don't meet these criteria are considered to be problematic. $^\&$A large fraction of the problematic transcripts for Oases and Trinity are likely caused by the presence of contaminated sequences or other artifacts so should not be considered as mis-assemblies. For example, 3,494 (out of 4,391) Oases transcripts have no significant alignments with the reference genomes with E-values better than 1e-4, and therefore are unlikely transcripts from the reference genomes. }
\end{table*}

\subsubsection{Accuracy evaluation for the TAG transcripts}
We blasted transcripts assembled by TAG against the three reference genomes to evaluate the accuracy of metatranscriptome assembly. Our results showed that only 16 out of 9,428 ($0.17\%$) transcripts cannot be perfectly aligned back to the reference genomes: among the 16 transcripts, 14 can be aligned with minor differences, and only two contain potentially serious problems (see Table 1). We note that there are two types of potential errors in the transcripts assembled by TAG: the errors introduced by TAG, and the errors inherited from the metagenome assembly (i.e., the mis-assemblies present in the metagenome assembly that propagates into the transcript). One of the problematic transcript is single-edge transcript, suggesting that this assembly error was propagated from the metagenome assembly. The other problematic transcript (of 390 bps) is a multi-edge spanning transcript, and the error was introduced by TAG (as no matching sequence can be found in the metagenome assembly).  Our results suggest that TAG achieves high assembly accuracy overall with an error rate of $<<1\%$.  If we only focused on multi-edge spanning transcripts (which are more difficult to assemble than transcripts contained within edges and therefore more error prone), the assembly error rate is still very low: only one out of 138 multi-edge transcripts contains such large assembly problem (the error rate is $0.7\%$).  

\subsubsection{Comparison with \emph{de novo} assembly}
We further compared the performance of TAG with Oases (\citealp{oases}) and Trinity (\citealp{trinity}) \footnote{Trinity has been applied to analyze metatranscriptomic datasets (\citealp{Microbiome2014}), although the program was developed targeting splicing isoforms in Eukaryotes}, \emph{de novo} assemblers for transcriptomic sequences. For Oases, we used merged results from assemblies using kmer sizes ranging from 19 to 31. Table 1 summarizes the comparison results. Although Oases and Trinity produced larger numbers of transcripts than TAG, the total bases in the transcripts assembled by these three methods are comparable (i.e., TAG assembled longer transcripts). If we considered only the ``good" transcripts by excluding the transcripts that cannot be aligned well to the reference genomes (which are likely misassembles, or assemblies from contaminated sequences or other artifacts commonly found in RNA-seq experiments (\citealp{biasRNAseq})), the difference in the total lengths of transcripts is even more significant. TAG produced a total of 9,426 good transcripts with a total of 7,020,484 bps, while Oases and Trinity assembled transcripts of 4,076,48 and 5,025,072 total bases, respectively. This result shows that using reference genomes for metatranscriptome assembly helps to improve the coverage and quality of the assemblies. 

We ran CD-HIT-EST (\citealp{cdhit}) to cluster the good transcripts from all programs at 95\% sequence identity cutoff, resulting in 10,944 clusters: only a modest number of clusters (2,309) are shared by all methods, 2,965 clusters are shared by two methods (1,399 shared Trinity and Oases; 1,369 by Trinity and TAG; and 116 by TAG and Oases ), and the remaining clusters are unique to one method (TAG: 2,571, Trinity: 2,983, and Oases: 197). This result suggests that de novo assembly and reference-based approaches can complement each other: transcripts of highly expressed genes in rare species (and therefore less well represented in metagenomes) may be assembled by de novo assembly, while transcripts of low expression level can only be identified using reference-based analysis. 

\subsection{Application of TAG to a real metatranscriptomic dataset}
We applied TAG to analyzing a metatranscriptomic dataset derived from a human stool sample, using its matched metagenomic dataset as the reference (\citealp{Giannoukos12}) \footnote{We combined the metatranscriptomic reads from four fractions of sequencing of the same sample, downloaded from SRA (SRX130930, SRX130937, SRX130922 and SRX130928), and the metagenomic reads from four fractions of sequencing, also downloaded from SRA (SRX130930, SRX130954, SRX130936 and SRX130949). Note that we used the metatranscriptomic dataset sequenced on 5 $\mu g$ RNA extracted from an individual's stool microbiome, which was shown to yield the best sequencing results (\citealp{Giannoukos12}).}. As described above, the metagenomic sequences were first assembled using SOAPdenovo2, and the metagenome assembly was then used as the reference for the metatranscriptome assembly by TAG. 

\subsubsection{Speed of the reads mapping to the de Bruijn graph}
Metatranscriptome assembly by TAG (including reads mapping onto the graph and the transcript inference afterwards) for this dataset takes about seven minutes to complete on a linux computer with Intel(R) Xeon(R) CPU E5-2680 v2 @ 2.80GHz (using single processor). The actual reads mapping step takes about one minute to complete---the remaining six minutes were spent on other I/O steps including processing the input SAM alignment file (from Bowtie2) and reads files. This indicates that our graph-based reads mapping algorithm (read2graph) is efficient. TAG adds only a small amount of computational time to the whole pipeline for the metatranscriptome analysis---SOAPdenovo2 takes several hours to assemble the metagenome, and mapping metatranscriptomic reads onto the metagenome contigs by Bowtie2 takes about 1,700 CPU minutes (the actual job was done in parallel using 32 processors). 

\subsubsection{Exploiting tangles in de Bruijn graph to improve metatranscriptome assembly}
We tested the performance of TAG using reference metagenomes assembled with different k-mer sizes, considering that the choice of k-mer size is important for the metagenome assembly (\citealp{soap,velvet}) and therefore metatranscriptome assembly. As shown in Figure 3, when a relatively small k-mer (e.g., 25) was used, the metagenome assemblies are more tangled, and as a result, fewer transcripts can be assembled using the contigs as the reference. This pitfall, however, can be alleviated by retaining the tangled structure (i.e., the ambiguous connection caused by short repeats) in the metagenome assembly in the de Bruijn graph, which can be exploited by TAG to connect metatranscriptomic reads into complete transcripts, resulting in improved assembly of metatranscriptome.  

As shown in Figure 3, the total length of the assembled transcripts by TAG decreases slowly when k-mer size increases from 25 to 31. Considering that most transcripts are longer at k-mer=31 as compared to smaller k-mers (e.g., average lengths of the transcripts are 264 and 273 for k-mer=25 and 31, respectively), we selected k-mer=31 to demonstrate the improvement of metatranscriptome assembly by using TAG.  Figure 4 shows the distribution of the path lengths (i.e., the number of edges that are traversed in the de Bruijn graph to form a transcript by TAG) of the transcripts assembled by TAG: most of the multi-edge transcripts span two edges (contigs), although a single transcript may span as many as seven edges. 

\begin{figure}[htbp]
   \centering
   \includegraphics[trim = 0mm 0mm 0mm 0mm, clip, scale=0.6]{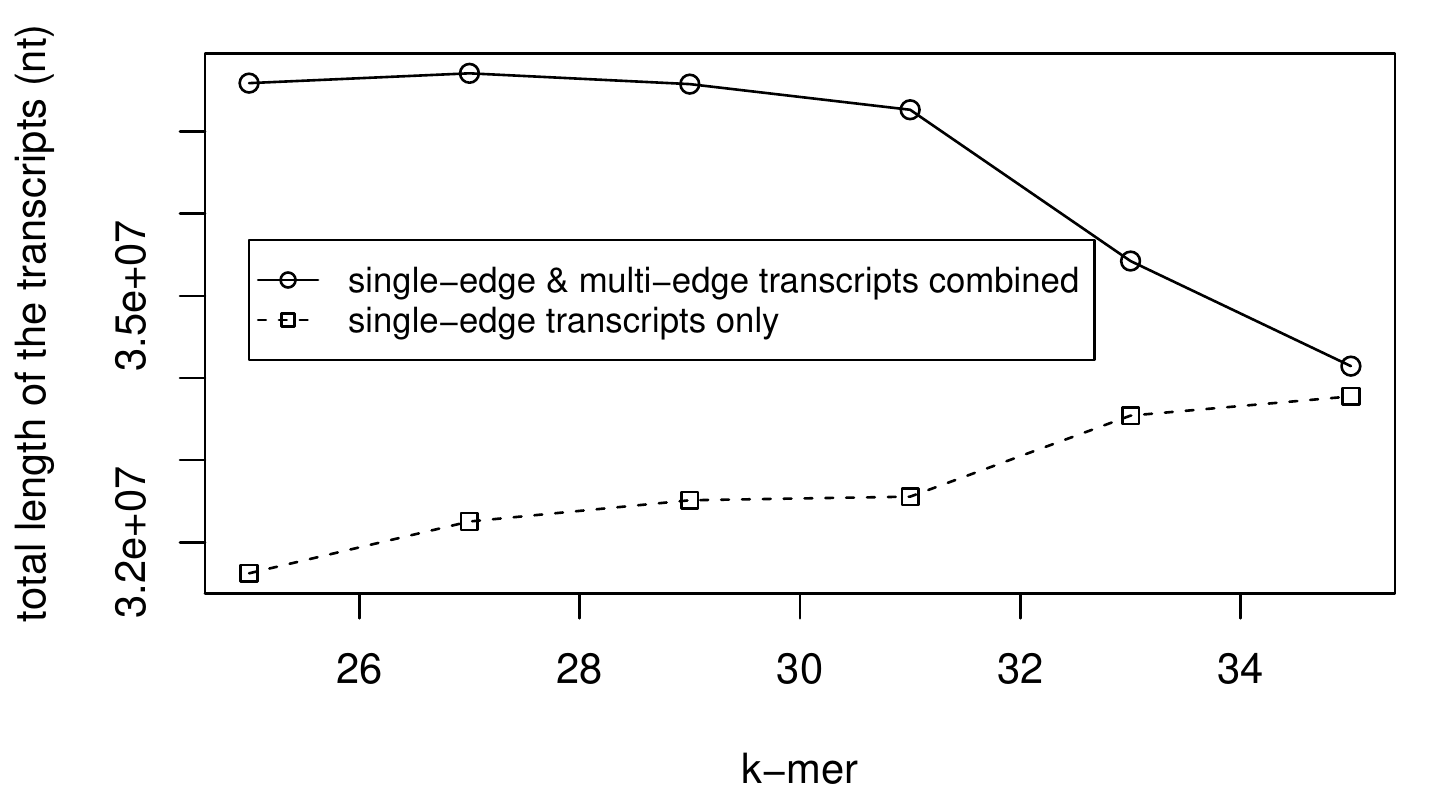}
   \caption{ The impact of k-mer size on the performance of TAG. When the k-mer size increases from 25 to 31 in SOAPdenovo2 assembly, the performance of TAG remains the same: a substantial fraction of multi-edge transcripts can be assembled by TAG. However, when further increasing the k-mer size to 35, most transcripts assembled by TAG are single-edge transcripts, indicating the TAG algorithm is not effective when a large k-mer is used. This is probably because, in this case, the metagenome assembly is fragmented rather than tangled, and as a result the total length of the transcript also decreases. Therefore, in the experiments of this paper, we choose k=31 in SOAPdenovo2 assembly, which seems to yield the best results here.}
   \label{fig:hash}
\end{figure}

\begin{figure}[htbp]
   \centering
   \includegraphics[trim = 0mm 0mm 0mm 0mm, clip, scale=0.6]{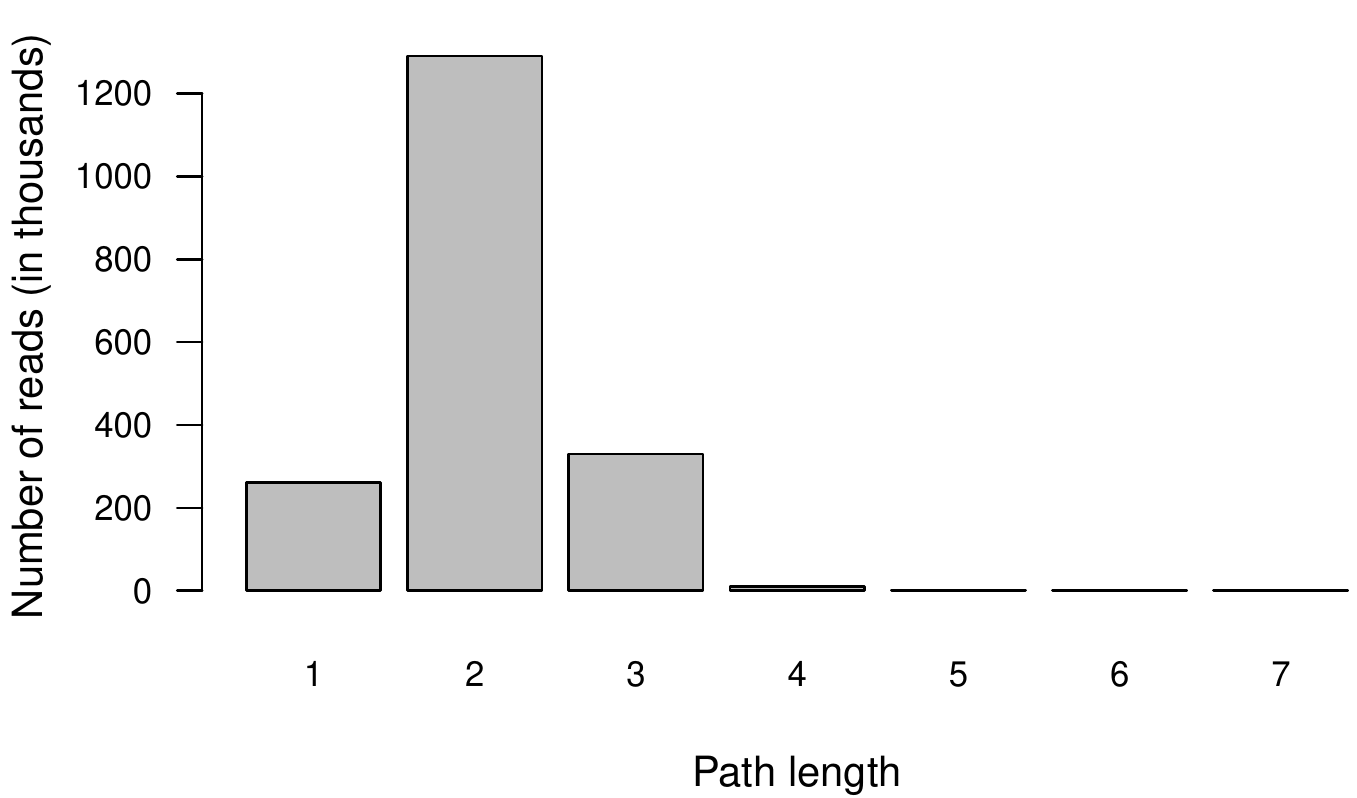}
   \caption{The path length distribution for {\em multi-edge-spanning reads} that span two or more edges when mapped to the de Bruijn graph by TAG. The X-axis represents the length of multi-edge-spanning read paths (i.e., the number of edges that the multi-edge-spanning reads span) and the Y-axis represents the total number of multi-edge-spanning reads spanning the paths of certain lengths.  Paths of length 1 represent the cases when the seed extension in one direction resulted in an alignment of at most 7 bps, and thus were considered insignificant and discarded. }
   \label{fig:hash}
\end{figure}

\begin{table*}[!t]
\caption{Some statistics of TAG assembly on the human stool metatranscriptomics dataset.}
\begin{center}
\renewcommand{\arraystretch}{1.4}
\setlength\tabcolsep{3pt}
\begin{tabular}{lll}
\hline
\noalign{\smallskip}
Total no. of reads & 27,962,127 x 2 (paired)\\
No. of reads mapped to contigs &19,233,474 x 2 +  7,645,742 (single) \\
No. of multi-edge-spanning reads & 1,893,157\\
No. of {\em resolved} $^*$ single-edge transcripts (length) & 112,527 (32,216,351 bps)\\
No. of {\em partial} $^*$ single-edge transcripts (length) & 2,573 (340,276 bps)\\
Total no. of {\em single-edge}$^\#$ transcripts (length) & 115,100 (32,556,627 bps)\\ 
No. of {\em resolved} of multi-edge transcripts (length) & 20,903 (4,596,622 bps)\\
No. of {\em partial} multi-edge transcripts (length) & 552 (110,063 bps)\\
Total no. of \emph{multi-edge}$^\#$ (length) transcripts (length) & 21,455 (4,706,685 bps)\\
Total no. of transcripts (length) & 177,463 (40,456,052 bps)\\
Proportion of multi-edge transcripts (in length) & 15.7\% (11.6\%)\\
\hline
\end{tabular}
\end{center}
{\footnotesize Only transcripts of at least 100 bps were considered in this summary. $^\#$Single-edge transcripts: the transcripts reported by TAG that are fully contained within edges (contig) in the de Bruijn graph of the metagenome assembly (they can be considered as the results of a baseline reference-based metatranscriptome assembly approach that uses the contigs as the reference);
Multi-edge transcripts: the transcripts reported by TAG that span multiple edges in the de Bruijn graph. $^*$
Partial transcripts: the transcripts that are not fully resolved by TAG (i.e., the edge sequences);
Resolved transcripts: the transcripts that are resolved by TAG and therefore likely represent full-length transcripts. }
\end{table*}

Table 2 summarizes the metatranscriptome assembly results by TAG. A majority of the metagenomic reads can be mapped to metagenomic assembly: for $68.8\%$ of read pairs, both reads can be mapped to contigs by Bowtie2, whereas  an additional $13.6\%$ reads can be mapped to contigs although their mate-pairs cannot be mapped. Among the $\approx 9.8 M$ remaining unmapped reads, $\approx 1.9M$ ($18.9\%$) can be mapped to multiple edges (i.e., through one or more junction k-mers in the de Bruijn graph) by TAG. Thanks to these reads, TAG was able to improve the metagenomic assembly significantly. In total, TAG assembled about 177K transcripts, among which about 21K ($15.7\%$) are multi-edge transcripts. These multi-edge transcripts cannot be fully assembled if only those reads mapped to contigs are considered in the metatranscriptome assembly; instead, they are likely to be broken into {\em partial} transcripts, each contained in a separate contig (i.e., the edge in the de Bruijn graph). We note that TAG did not resolve all transcripts. A small fraction of TAG-assembled transcripts are {\em partial} transcripts, each of which represents a unique edge in the tangled {\em transcript graph}, formed by two or more transcripts sharing some common segments (see Methods for details) that cannot be resolved without additional information. About $2.6\%$ (552 out of 21,455) of the multi-edge transcripts were not fully resolved by TAG and remained as partial transcripts. Similarly, $2.2\%$ (2,573 out of 115,100) of the single-edge transcripts are also partial transcript as some multi-edge-spanning reads connect them with other partial transcripts, although their actual connections remain ambiguous. We note that these two numbers increase substantially (to 21.1\% and 8.1\%, respectively)  when there is no minimum length applied for output transcripts.

\section{Discussion}

Even though thousands of complete prokaryotic genomes and many more draft genomes are available, metagenomes are constantly found to contain many new species and new genes (\citealp{Vital14,gut10,hmp12}). It is therefore important to develop methodologies for metatranscriptome data analysis that are not constrained by the sequenced genomes. With ``matched" metagenomic and metatranscriptomic datasets, we believe that proper utilization of the metagenome data will help greatly the analysis of metatranscriptomic data (and vice versa). The eventual integration of these datasets (as well as other meta-omic datasets) will provide new insights on the composition, function, and regulation of microbiomes. Well assembled transcripts are important for the function annotation of the metatranscriptome, and also for inferring gene regulatory mechanisms such as the operons.  

We developed a novel reads mapping algorithm (read2graph) that allows fast mapping of short reads from transcriptome sequencing onto the assembly graphs of reference genomes. We applied this mapping algorithm for metatranscriptome assembly, showing the utility of the de Bruijn assembly graph of the metagenome in downstream applications such as the metatranscriptome analysis. Our mapping tool is fast and can be applied to other applications, for example, mapping metagenomic sequencing reads onto the de Bruijn graph of closely related species for estimating the relative abundances of these species (\citealp{Wang12}).  We have shown in a related research that genes are often broken into fragments in metagenome assembly, and multi-edge-spanning reads can stitch them together (\citealp{genestitch}). The mapping of multi-edge-spanning reads will also improve quantification of gene expression based on read counts, in particular for genes (from the same or different organisms) sharing highly similar sequences. In reality, however, we may still miss the mapping of a small fraction of multi-edge-spanning reads: if a read contains a sequencing error in the occurrence of a branching k-mer, we cannot find its location in the graph. Because of the low error rate ($<$1\%) in Illumina reads, we believe this fraction of reads is indeed negligible in metatranscriptomic data analysis. 

We note that de Bruijn graphs will naturally capture the genomic variations of the metagenomes in the graphs (e.g., the single-nucleotide variations are represented as bulges (\citealp{marygold}), the variations in tandem repeats are represented as wheels, and structural variations are represented long loops (\citealp{euler01})), which is yet another advantage of using graphs instead of contigs to represent metagenomes. Genomic variations in metagenomes are naturally handled by our graph-centric mapping approach. 

We expect that a combination of different approaches (reference-based and \emph{de novo}) need to be applied to accomplish the comprehensive metatranscriptome analysis. As the references for metatranscriptome analysis, the matched metagenome will never be perfect, due to biological (rare species may be poorly sampled), experimental (some genomic regions may not be covered well), and computational (assemblers are not perfect) reasons. Integration of known reference genomes, matched metagenomes, and even non-matched metagenomes can maximize the coverage of references for reference-based approaches. On the other hand, if a microbial community contains new, rare but highly expressed microbial species, their transcripts can only be revealed by de novo metatranscriptome assembly (\citealp{Schulz2012}) but not by the reference-based approaches such as the one presented in this paper.  

\section*{Acknowledgements}
\paragraph{Funding\textcolon} This research was supported by NIH grant 1R01AI108888-01A1.

\bibliographystyle{natbib}

\end{document}